\documentclass{aa}  
\usepackage[switch]{lineno}
\usepackage{graphicx}
\usepackage{txfonts}
\usepackage{multirow}
\usepackage{booktabs}
\usepackage{enumitem}
\usepackage[colorlinks=true, linkcolor=blue, citecolor=blue, urlcolor=blue]{hyperref}
\begin{document}

   \title{Broadband Modelling of GRB~230812B Afterglow: Implications for VHE $\gamma$-ray Detection with IACTs}

   \subtitle{}

   \author{
Shraddha~Mohnani\thanks{\email{\href{mailto:mohnani.shraddha97@gmail.com}
{mohnani.shraddha97@gmail.com}}}$^{\inst{\ref{ins1},\ref{ins2}}}$,
Biswajit~Banerjee\thanks{\email{\href{mailto:biswajit.banerjee@gssi.it}{biswajit.banerjee@gssi.it}}}$^{\inst{\ref{ins2}, \ref{ins3}}}$
\and Davide~Miceli$^{\inst{\ref{ins4}}}$
 \and Lara~Nava$^{\inst{\ref{ins5}, \ref{ins6}}}$
 \and Gor~Oganesyan$^{\inst{\ref{ins2},\ref{ins3}}}$
 \and Pawan~Tiwari$^{\inst{\ref{ins2},\ref{ins3}}}$
 \and Annarita~Ierardi$^{\inst{\ref{ins2},\ref{ins3}}}$
 \and Alessio~L.~De Santis$^{\inst{\ref{ins2}, \ref{ins3}}}$
\and Samanta~Macera$^{\inst{\ref{ins2}, \ref{ins3}}}$
\and Amit~Shukla$^{\inst{\ref{ins1}}}$ 
 \and Marica~Branchesi$^{\inst{\ref{ins2}, \ref{ins3}}}$
 \and Swarna~Chatterjee$^{\inst{\ref{ins1}, \ref{ins7}}}$
 \and Sushmita~Agarwal$^{\inst{\ref{ins1}}}$
 \and Abhirup~Datta$^{\inst{\ref{ins1}}}$
 \and Kuldeep Kumar Yadav$^{\inst{\ref{ins9}}}$
 \and G.C. Anupama$^{\inst{\ref{ins10}}}$
 }
 \institute{Department of Astronomy, Astrophysics and Space Engineering Indian Institute of Technology Indore, Simrol, Khandwa Road, Indore 453552, Madhya Pradesh, India. \label{ins1} 
 \and
 Gran Sasso Science Institute, Viale F. Crispi 7, L'Aquila (AQ), I-67100, Italy. \label{ins2}
\and
INFN - Laboratori Nazionali del Gran Sasso, L'Aquila (AQ), I-67100, Italy.
\label{ins3}
\and
INFN - Sezione di Padova I-35131 Padova Italy
\label{ins4}
\and
INAF - Osservatorio Astronomica di Brera, Via E. Bianchi 46, I-23807, Merate (LC), Italy
\label{ins5}
\and
INFN, Sezione di Trieste, I-34127 Trieste, Italy
\label{ins6}
\and
Centre for Radio Astronomy Techniques and Technologies, Department of Physics and Electronics, Rhodes University, Makhanda 6139, South Africa
\label{ins7}
\and
Astrophysical Sciences Division, Bhabha Atomic Research Centre, Mumbai 400085, India
\label{ins9}
\and
Indian Institute of Astrophysics, II Block, Koramangala, Bengaluru 560034, India
\label{ins10}
}
\authorrunning{Mohnani et. al 2025}
   \date{}

  \abstract
  {A significant fraction of the energy from the $\gamma$-ray burst (GRB) jets, after powering the keV-MeV emission, forms an ultra-relativistic shock propagating into the circumburst medium. The particles in the medium accelerate through the shock and produce afterglow emission. Recently, few GRB afterglows have been observed in TeV $\gamma$-rays by Cherenkov Telescopes. This provides access to broadband spectra of GRB afterglows containing rich information about the microphysics of relativistic shocks and the profile of the circumburst medium.
  Since the transition from synchrotron to inverse Compton (IC) regime in afterglow spectra occurs between hard X-rays and the very-high-energy (VHE) $\gamma-$rays, detection in one of these bands is required to identify the two spectral components. The early afterglow data in the hard X-rays, along with the GeV emission, could accurately constrain the spectral shape and help in capturing the spectral turnover to distinguish the two components. We present the multiwavelength spectral and temporal study, focused on the keV-VHE domain, of GRB~230812B, one of the brightest GRBs detected by Fermi Gamma Ray Burst Monitor (\textit{Fermi}/GBM), along with the detection of a 72 GeV photon in Large Area Telescope (\textit{Fermi}/LAT) during the early afterglow phase. Through a detailed modeling of the emission within the afterglow external forward shock in a wind-like scenario, we predict up to $\sim1$ day optical to high-energy observations. We emphasize the importance of following up poorly localised GRBs by demonstrating that even in cases without prompt sub-degree localisation, such as GRB~230812B, it is possible to recover the emission using imaging atmospheric Cherenkov telescopes (IACTs), thanks to their relatively wider field of view. Moreover, we show that the low energy threshold of Large-Sized Telescope (LST) is essential in discovering the VHE component at much higher redshifts, typical to long GRBs.
}
  
   \maketitle

\section{Introduction} \label{sec:intro}
$\gamma$-ray bursts (GRBs) are fast MeV transients from ultra-relativistic jets produced after the collapse of massive stars  or compact binary mergers. These jets are powered by a central engine, typically a newly formed black hole \citep{2001ApJ...557..949N} created in such cataclysmic events. Internal dissipation of energy \citep{1994ApJ...430L..93R, 2000A&A...358.1157D} within the jet gives rise to highly variable (0.1-1s) keV–MeV radiation, observed as the prompt emission. The residual energy of the jet is then transferred to the circumburst medium, through the formation of a relativistic collisionless shock. The particles from the medium accelerate through the shock and emit non-thermal radiation usually by synchrotron and self-synchrotron losses resulting into a broadband afterglow spanning $\gamma$-rays to radio waves \citep{1997ApJ...476..232M, 1993ApJ...418L...5P, Sari_Piran_Narayan1997}. 

Since the launch of Fermi Space Telescope, thousands of $\gamma$-ray bursts have been detected by the Gamma Ray Burst Monitor \citep[\textit{Fermi}/GBM;][]{Fermi_GBM} in the energy range 8\,keV-40\,MeV, and around $10\%$ of them have been captured by the Fermi Large Area Telescope \citep[\textit{Fermi}/LAT;][]{Fermi_LAT} in 0.03-300\,GeV \citep{4FGGC,3FGGC, 2FGLC}\footnote{\url{https://fermi.gsfc.nasa.gov/ssc/observations/types/grbs/}}. The afterglow emission observed by \textit{Fermi/LAT} has been widely understood to be originating from the external forward shock via synchrotron emission \citep{Kumar_2010,Ghisellini_2010,Nava_2011}. However, the GRBs detected at very-high-energy $\gamma$-rays (VHE; $\rm{E>}$100\,GeV) are only a handful. The discovery of GRB~190114C \citep{TeV_in_190114C} and GRB~180720B \citep{GRB180720B} unequivocally proved for the first time that GRB afterglows can produce photons above 100\,GeV. Later, a close-by $\gamma$-ray burst at z=0.078, GRB~190829A \citep{GRB190829A} showed that VHE emissions from GRBs can last up to long timescales ($\sim$days). 
In addition GRB~201015A and GRB~201216C also established as TeV emitters \citep{GRB201015A,GRB201216C}, where the latter was the farthest GRB detected so far with a redshift of 1.1 \citep{GRB201216C}.
The detection of the GRBs starting from the trigger time using imaging atmospheric Cherenkov telescopes (IACTs) is challenging due to the communication of the trigger of a burst from a satellite (\textit{Fermi} and/or \textit{Swift}) to the IACTs and the slew-time (typically around 20-30\,s) required to start the observation. In contrast to this, GRB 221009A, the brightest $\gamma$-ray burst of all times  was observed since the time of the trigger by LHAASO \citep{BOAT_LHAASO} until 6\,ks thanks to the larger field of view (FoV) of the telescope detecting gamma-ray photons up to 10 TeV. 

The broadband GRB afterglow emission from radio to high-energy is typically attributed to synchrotron radiation from particles accelerated across the shock produced by the blast wave as it propagates into the medium \citep{sari_1996,Sari_Piran_Narayan1997, Granot_1999, Panaitescu_and_Kumar_2001, Zhang_2006}. Several works have theoretically explored the presence of another emission component in GRB spectra arising from synchrotron self-Compton (SSC) scattering of afterglow photons from the  electrons \citep{Sari_Esin_2001,1996ApJ...471L..91P, Derishev_piran_2021}. The availability of rich afterglow data from X-rays to TeV has made it possible to explore advanced models, such as inhomogenous magnetic fields in the shocks \citep{Khangulyan_Aharonian_et_al} or pair loading effects \citep{pair_loading}, or early inverse jet breaks \citep{Derishev_Piran221009A}. The detection of a second component has been claimed in TeV emissions from GRB190114C and GRB180720B as well \citep{TeV_in_190114C, GRB180720B}. However, in the case of GRB190829A, there is ambiguity, since the joint spectra can be explained by a single-component emission, while other studies \citep{2022ApJ...931L..19S} indicate that a two-component model is required. 

The sub-MeV to GeV observations during the early afterglow are essential to mark the end of the first high-energy component and the rise of the second component, commonly modelled with SSC. However, capturing the sub-MeV emission from slowly varying transients is challenging, as MeV instruments are typically background dominated. Although Swift's Burst Alert Telescope \citep[\textit{Swift}/BAT;][]{Swift/BAT} is sensitive in the hard X-ray band up to 150 keV, it detects about a factor of three less GRBs per year as compared to \textit{Fermi}/GBM. Hence, in case of bright GRBs, \textit{Fermi}/GBM data can be leveraged to extract the sub-MeV afterglow using the orbital background subtraction method \citep{osv_tool}. Such an early afterglow has previously been detected and studied in the case of  GRB~221009A by \cite{Banerjee.et.al}.

Triggered on 18:58:12 UT, August 12, 2023 ($\rm{T_0}$), GRB~230812B, is one of the brightest $\gamma$-ray bursts detected by \textit{Fermi}/GBM, with a fluence of $2.52\rm{\times 10^{-4} erg/cm^2}$ \citep{GCN34391}. The duration ($T_{90}$)  as recorded by \textit{Fermi}/GBM is 3.26 s \footnote{\url{https://heasarc.gsfc.nasa.gov/FTP/fermi/data/gbm/bursts/2023/bn230812790/current/}glg\_bcat\_all\_bn230812790/\_v02.fit}. High photon flux from the burst caused pulse pileup in GBM detectors for $\sim$1\,s \citep{GCN35660}. The burst has also been detected in \textit{Fermi}/LAT from the onset of prompt 
\citep{GCN34392} up to about 1000\,s. The highest-energy photon, with an energy of 72~GeV, was detected by LAT at approximately 32~s after the trigger.
 Since \textit{Swift}/BAT did not detect the prompt emission, the burst could only be broadly localized by \textit{Fermi}/GBM in the initial phase. After a delay of $\sim$25\,ks from the GBM trigger, the Swift X-ray Telescope \citep[\textit{Swift}/XRT;][]{Swift/XRT} detected the X-ray afterglow and precisely localised the burst at R.A. = 249.13$^\circ$ Dec = +47.86$^\circ$ (J2000) \citep{GCN34400}. Later on, the optical observations from several telescopes revealed a presence of supernova associated with the burst \citep{GCN24500} confirming the stellar collapse origin. The spectroscopic observations from GTC and NOT optical telescopes measured the redshift of $z=0.36$ \citep{GCN34409, GCN34410}. Subsequently the radio afterglow was first detected by Arcminute Microkelvin Imager Large-Array (AMI-LA) around 2 days post burst at 15.5 GHz with a flux 280 $\rm{\mu}$Jy  \citep{GCN34433}. Around the same time ($\sim$2.3 days), it was also followed up by Karl G. Jansky Very Large Array (VLA)  at 6 Hz and 10 GHz detecting a flux of 230 $\rm{\mu}$Jy and 196 $\rm{\mu}$Jy respectively \citep{GCN34552}. However, in further observations from $\sim$17 days post-burst onward, the source was not detectable in the radio sky, and only upper limits were obtained by VLA \citep{GCN34735} and upgraded Giant Metrewave Radio Telescope (uGMRT) \citep{GCN34727}.

The extreme brightness of the GRB~230812B and discovery of the 72 GeV photon motivated us to perform a broadband spectral study of the burst from prompt to afterglow phase.  Our primary aim is to study the joint MeV–GeV emission in the early afterglow, along with the broadband emission at later times, to infer the microphysical properties of the burst.  This would also help in understanding the spectral component(s) giving rise to GeV emissions. Furthermore, we explore the possibility of VHE emission from the burst, detectable by existing facilities.

This paper is structured as follows. In Sect.~\ref{sec:MWLdata}, we describe the methodology adopted for the analysis of multiwavelength data obtained from various telescopes. Sect.~\ref{results} describes the results of our spectral analysis. We report the detection of rare MeV afterglow. In Sect.~\ref{sec:spec_model} we describe the spectral and temporal modeling of the burst and discuss a possible strategy to effectively capture the VHE emissions from such GRBs. Finally in Sect.~\ref{discussion} we summarize the complete evolution of the multiwavelength spectra and the microphysics of the emission region.

\section{Multiwavelength data analysis} \label{sec:MWLdata}
We performed a multiwavelength analysis of GRB~230812B to study its spectral and temporal evolution from prompt to afterglow phase.  We analysed the high-energy $\gamma$-ray (HE; 0.1$\,<\,$E$\,<\,$100\,GeV) from \textit{Fermi}/LAT, hard X-ray data from \textit{Fermi}/GBM in 8\,keV - 40\,MeV, late time X-ray data from \textit{Swift}/XRT in 0.3 - 10\,keV, and radio data in 1.4 GHz from uGMRT. 
The data in the optical r$^{\prime}$ band has been collected from \cite{SN_paper_GRB230812B}.
In the following sections, the multiwavelength data analysis methods (or sources from which data have been adapted) are described in detail. 

\subsection{\textit{Fermi}/GBM}\label{sec:gbm}
We divided the prompt emission phase ($\rm{T_{90}}$) into three temporal bins: 0–0.4\,s (Bin-1), 1.4–2.0 s (Bin-2), and 2.0–3.6 s (Bin-3).  \textit{Fermi}/GBM data from 0.4–1.4 s are excluded to avoid pulse pile-up and dead-time effects caused by the high photon flux, as recommended by the \textit{Fermi}/GBM team \citep{GCN35660}. The time interval following the end of $T_{90}$, i.e., 3.6–10.2 s (Bin-4), possibly marks the transition from the prompt to the afterglow phase. The afterglow is divided into three temporal bins: 10.2–25 s (Bin-5), 25–250 s (Bin-6) and 251.9-647.2 s (Bin-7). 

\subsubsection{Prompt emission}
 For the spectral analysis of prompt emission, we utilized \textit{Fermi}/GBM time tagged events (TTE) data, publicly available through the HEASARC GBM-burst catalog\footnote{\url{https://heasarc.gsfc.nasa.gov/FTP/fermi/data/gbm/bursts/}}. We rebin the TTE data to a resolution of 64 ms by \textit{binbytime} method using the binning module for unbinned data in \textit{Fermi}/GBM data tools \citep{GbmDataTools}. To remove the bad time intervals (BTI) in our analysis, we ignored the time bins consisting of photons that arrive between 0.4-1.4 s \citep{GCN35660}.  
 For background estimation, we selected intervals (-20, -5) and (50, 80) s with respect to the trigger time and fitted it with a polynomial function in GBM data tools. We found that the background can be fitted reasonably well with the first-order polynomial. The model is then interpolated in the source region and the modeled background counts are estimated for further analysis.
 
 Subsequently, the spectrum, background and response files have been extracted. We fit the prompt emission spectra using \texttt{HEASOFT XSPEC version: 12.15.0}. We used 4 NaI detectors n0, n3, n6, n7  in energy range 8-900 keV and one BGO detector b0 with energy range 0.32-40 MeV for the analysis. The choice of NaI detectors is based on pointing angle with the source to be less than 60$^{\circ}$. For BGO, we selected the detector with the minimum pointing angle. For the spectral fit of this dataset we used Poisson-Gaussian statistics (pgstat\footnote{\url{https://heasarc.gsfc.nasa.gov/docs/xanadu/xspec/manual/XSappendixStatistics.html}}) in \texttt{XSPEC} \citep{xspec}. The fit results are shown in table \ref{tab:GBM_results}.

\subsubsection{Afterglow}\label{GBM_afterglow}

To accurately estimate the background and detect faint MeV emission during the afterglow, we used the orbital background subtraction technique \citep{osv_tool,OSV_in_BATSE}, which takes advantage of the periodic observing geometry of the Fermi spacecraft. Calculating the average counts over the consecutive orbits enveloped in the desired time can effectively estimate the background. We used the daily \texttt{CSPEC} data\footnote{\url{https://heasarc.gsfc.nasa.gov/FTP/fermi/data/gbm/daily/2023/08/12/current/}} from the day of the GRB event to generate source and background files. We estimated the background over 30 orbits using the Fermi GBM Orbital Background Subtraction Tool\footnote{\url{https://fermi.gsfc.nasa.gov/ssc/data/analysis/user/}Fermi\_GBM\_OrbitalBackgroundTool.pdf} \citep[OSV;][]{osv_tool}. Since the time of interest lies outside the duration of the prompt emission, using the standard response files available in the \textit{Fermi}/GBM catalog is not suitable for the analysis and might introduce inaccuracies. To generate precise response matrices at the source location, for each detector of interest for each time bin, we used the official Fermi tool GBM Response Generator\footnote{\url{https://fermi.gsfc.nasa.gov/ssc/data/analysis/gbm/DOCUMENTATION.html}} on the daily data of the $\rm{\gamma}$-ray burst. After generating the spectrum, response and background files, we fit the spectra for Bin-5 and Bin-6 using Powerlaw (PL) function in \texttt{XSPEC}. We use the \texttt{XSPEC} model \texttt{cflux*powerlaw} in order to calculate the integrated flux from the fit. We frozen the normalization parameter of the \texttt{powerlaw} function to 1 since the model is now normalized by the integral flux. However, for Bin-4, we tested the spectral fit against the models PL, Smoothly Broken Powerlaw (SBPL), Band, Band + Powerlaw, and Broken Powerlaw (BPL). For the afterglow bins Bin-5 and Bin-6, the two NaI detectors n6 and n3 were selected as they have the lowest pointing angles with the source (28.2$^{\circ}$ and 24.6$^{\circ}$, respectively) compared to other detectors throughout the time of interest. We analysed the afterglow spectrum in time range 40-400 keV covering the most sensitive energy range of the NaI detectors \citep{osv_tool, 4FGGC, 2FGGC, GBM_instrumentation}. For GRB~230812B, we did not find any excess over the background in the BGO detector in the entire energy range 0.4-40 MeV during the afterglow bins (Bin5 and Bin6). Due to the low photon count, we use the cash statistic \citep{cash} as the fit statistic for our analysis.

\subsection{Fermi-LAT}\label{Fermi_LAT_analysis} 
We performed unbinned likelihood analysis of \textit{Fermi}/LAT data for the $\rm{\gamma}$-ray burst, starting from the \textit{Fermi}/GBM trigger time and extending up to 1000 seconds post-trigger, in the energy range of 0.1–100 GeV, using the \texttt{GTBURST} \footnote{\url{https://fermi.gsfc.nasa.gov/ssc/data/analysis/scitools/gtburst.html}} software from Fermi Science tool. We selected the region of interest covering $12^{\circ}$ around the source location R.A. = 249.13$^\circ$, Dec = +47.86$^\circ$ (J2000) given by \textit{Swift}/XRT observations \citep{GCN34400}. We used the \texttt{P8R3$\_$TRANSIENT020} event class, which is suitable for transient-source analysis, and the corresponding instrument response functions. Isotropic particle background (\texttt{`isotr template'} in \texttt{GTBURST}), galactic and extragalactic high-energy components from the Fermi Fourth Catalog (4FGL), with fixed normalization (\texttt{`template (fixed norm)'}) has been used. To ensure $5 \sigma$ detection, a non-uniform time binning scheme has been adopted with a criteria of Test-statistic $\rm{(TS) >25}$ to fit the power-law (\texttt{`powerlaw2'} using \texttt{GTBURST}) spectral model. For the time bin 250-647 s, we found the flux upper limits with a TS = 20. The time resolved spectral analysis as presented in table \ref{tab:LAT}. A 72 GeV photon at $\sim 32.13$~s has been detected \citep{GCN34392} with a 99.99\% probability of association with the GRB as found by employing the \texttt{gtsrcprob} method in \texttt{GTBURST}.

\subsection{Swift-XRT}
The X-ray afterglow of the $\rm{\gamma}$-ray burst has been observed by Swift-XRT starting approximately 25\,ks after the \textit{Fermi}/GBM trigger.
We obtained the source and background spectral files, as well as the redistribution matrix and ancillary response files, for the Swift XRT data in the 0.3–10 keV energy range from the online Swift-XRT GRB spectrum repository \citep{Swift-XRT_repository}. The spectra were extracted in photon counting mode for four time bins:  (25.4-27.2~ks), (30.9-38.2~ks), (191.7-215.6~ks) and (407.8-1466~ks). We fitted the XRT data for all four time bins simultaneously with an absorbed powerlaw model in \texttt{XSPEC}, using \texttt{tbabs*ztbabs*cflux*powerlaw}. Cflux is used to calculate the unabsorbed flux in the 0.3-10 keV energy range. The \texttt{XSPEC} models \texttt{tbabs} and \texttt{ztbabs} take into account the Tuebingen-Boulder ISM absorption in the Milky-Way and the host galaxy, respectively. The Hydrogen column density of the Milky-Way in the burst direction has been fixed to $\rm{N_H} =2.02\times10^{20}\,\rm{cm^{-2}}$ \citep{Willingale_2013}. The Hydrogen column density of the host galaxy at redshift z = 0.36, $N_H(z)$, is left as a free parameter for the fit, common among all the spectra. 

\subsection{Optical}

The optical afterglow and the associated supernova (SN) SN2023pel were monitored by several optical telescopes. We retrieved the r$^{\prime}$-band optical data from \cite{SN_paper_GRB230812B}, which were compiled from the GCN Circulars archive. The data set includes observations from various optical telescopes provided in Tab.~A2 in \cite{SN_paper_GRB230812B}.
We fitted the decaying part of the light curve, prior to the emergence of the supernova, to model the temporal evolution of the afterglow (see Sect.~\ref{sec:LC_model} for details) until the emergence of the SN. While observations in other optical filters were available, they were excluded from temporal fitting, as our aim is to model the burst's temporal behavior and jet dynamics rather than performing a detailed spectroscopic analysis for a band in which the source was seen at the earliest time and has a high cadence of observation. However the data from multiple filters were used to construct the spectral energy distribution (SED) around 30\,ks which was further used to test the spectral modeling (Figure \ref{fig:ssc_model}).

\begin{table*}[ht!]
    \centering
    \begin{tabular}{c c c c c c c c } \hline
    \multirow{2}{*}{Bin}  & {t-T$_{0}$} & {Model} & $\alpha$ & $\beta$ & $E_{peak}$ & Flux & stat/ dof \\ 
           & [s]        &       &        &  &   [keV] & [erg cm$^{-2}$ s$^{-1}$]& \\ \hline
    Bin-1   & 0-0.4     &  Band &  $-0.61\,\pm\,0.02$     & $-3.25\,\pm\,0.33$  & $813.1\,\pm\,58.7$ & $(1.47\,\pm\,0.06)\times10^{-4}$ & 758/561\\
    Bin-2   & 1.4-2.0     &  Band & $-0.22\,\pm\,0.03$     &  $-3.03\,\pm\,0.09$  & $77.9\,\pm\,2.3$ & $(5.67\,\pm\,0.13)\times10^{-5}$  & 699/561\\
    Bin-3   & 2.0-3.6     &  Band & $-0.65\,\pm\,0.03$ &  $-2.97\,\pm\,0.07$ & $63.9\,\pm\,2.3$ & $(1.70\,\pm\,0.04)\times10^{-5}$ &799/561\\ \hline
    Bin-4   & 3.6-10.2    &  BPL & $-1.66\,\pm\,0.04$ & $-3.00\,\pm\,0.06$  &$99.2\,\pm\,3.1$ & $(6.02 \pm 0.05) \times 10^{-6}$ & 805/135\\ 
    Bin-5   & 10.2-25.0   & PL    & $-2.30\pm 0.32$ & --& -- & $(8.12\pm1.66) \times 10^{-8}$  & 248/133\\ 
    Bin-6   & 25.0-250.0  & PL    &$-2.40\pm 0.23$&   --& --& $(2.30\pm 0.32) \times 10^{-9}$ & 200/135 \\
    Bin-7   & 251.9-647.2 & PL    &  $-1.41\pm 0.63$  & -- & --& $(6.43\pm 2.78) \times 10^{-10} $ & 194/133 \\ \hline
    \end{tabular}
    \caption{Time-resolved spectral analysis of GRB~230812B using \textit{Fermi}/GBM data. Bins 1--3 (from trigger to \( \rm{T_{90}} \)) represent the prompt emission, well-described by the Band function in the 8\,keV--40\,MeV range. Bin 4 marks the transition phase from prompt to afterglow, while Bins 5--7 correspond to the keV-MeV afterglow, fitted with a power-law model in the 40--400\,keV range. The best-fit spectral models for each time bin are listed in column 3.}

    \label{tab:GBM_results}
\end{table*}
\begin{figure*}[ht!]
    \includegraphics[width=0.33\textwidth,]{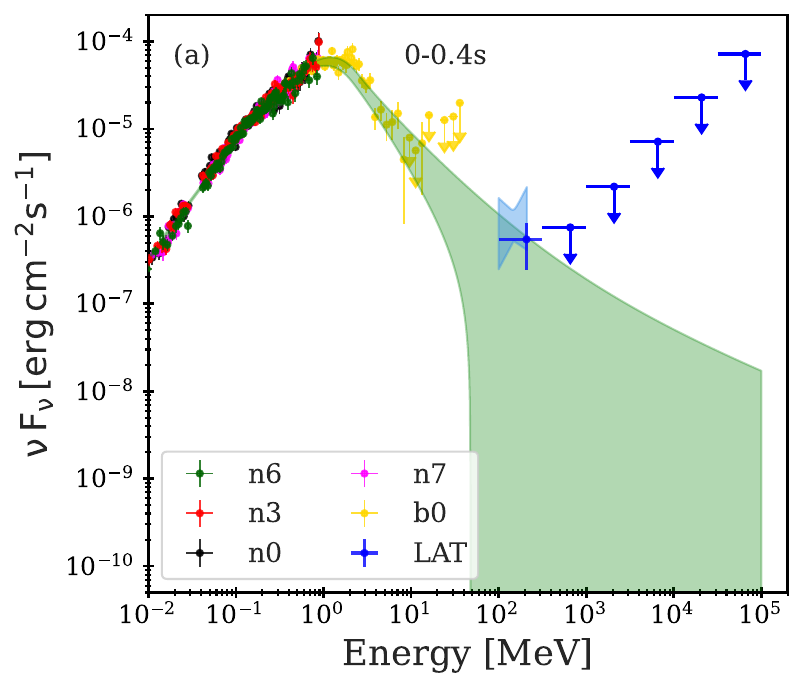}
    \includegraphics[width=0.33\textwidth,]{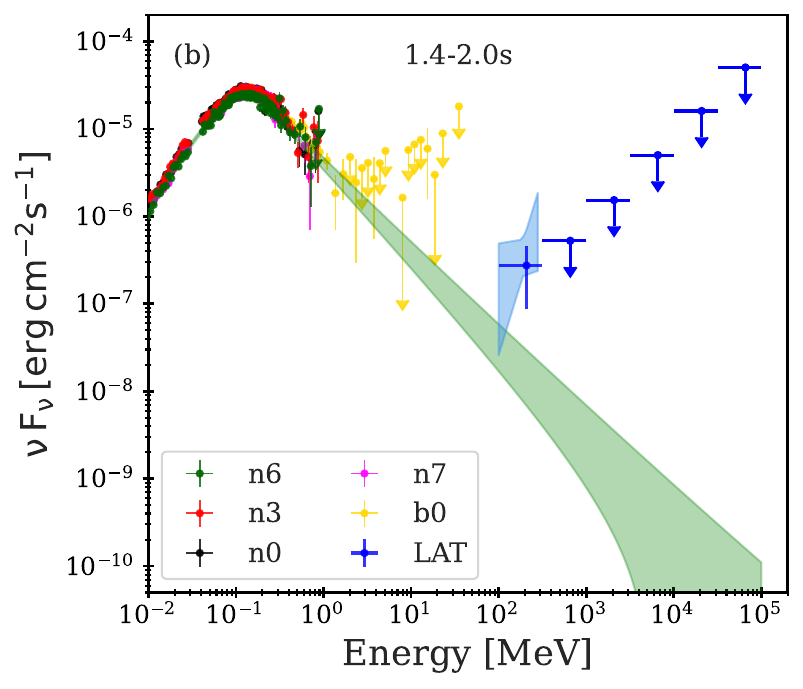}
    \includegraphics[width=0.33\textwidth,]{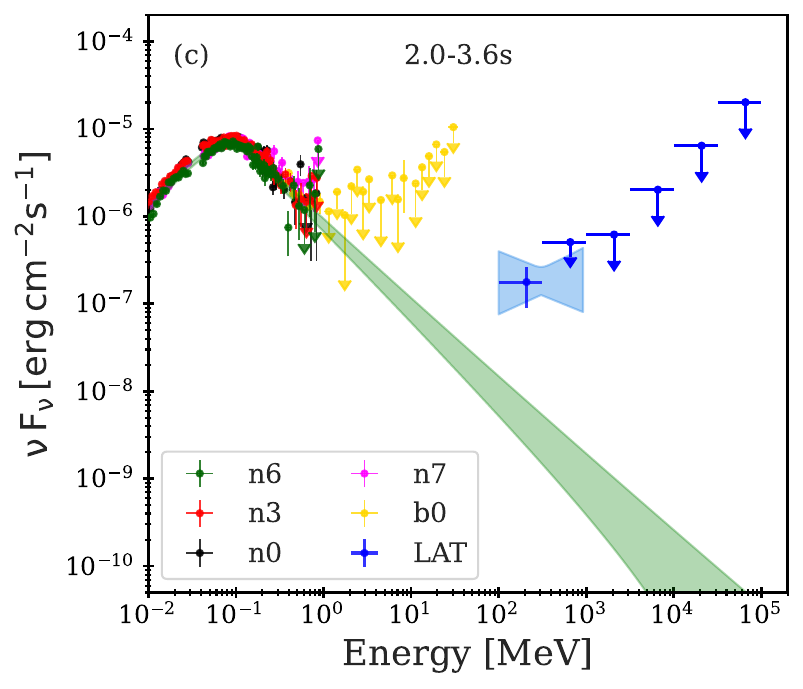}
    \includegraphics[width=0.33\textwidth,]{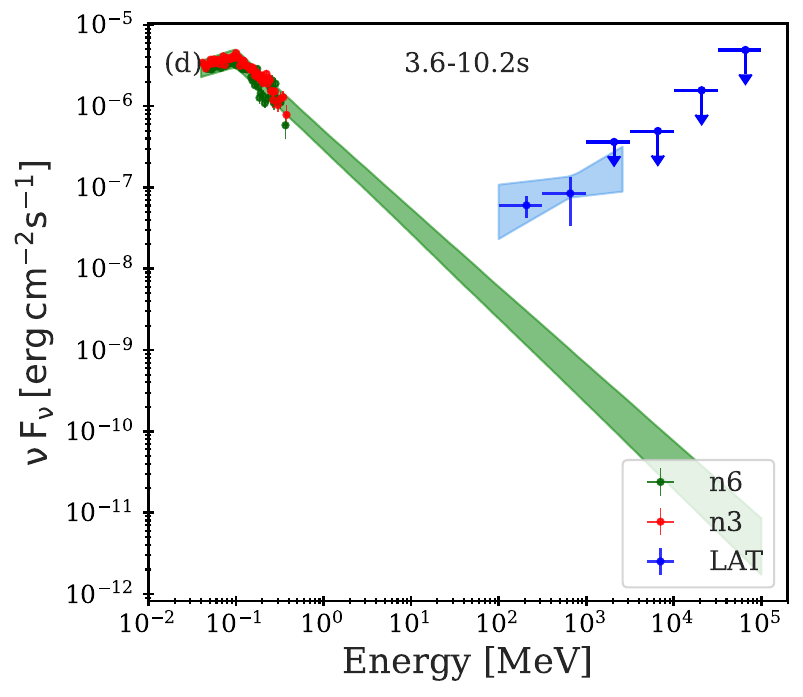}
    \includegraphics[width=0.33\textwidth,]{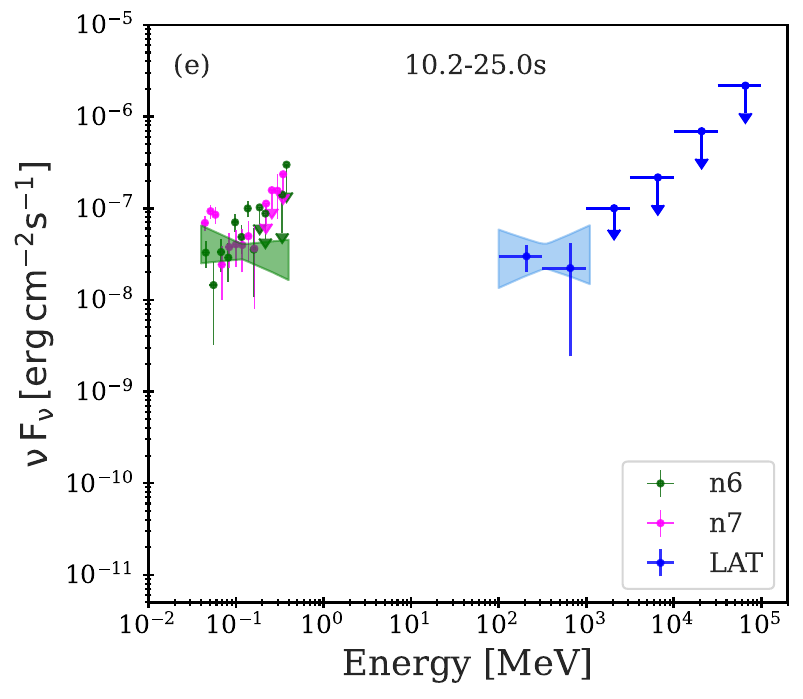}
    \includegraphics[width=0.33\textwidth,]{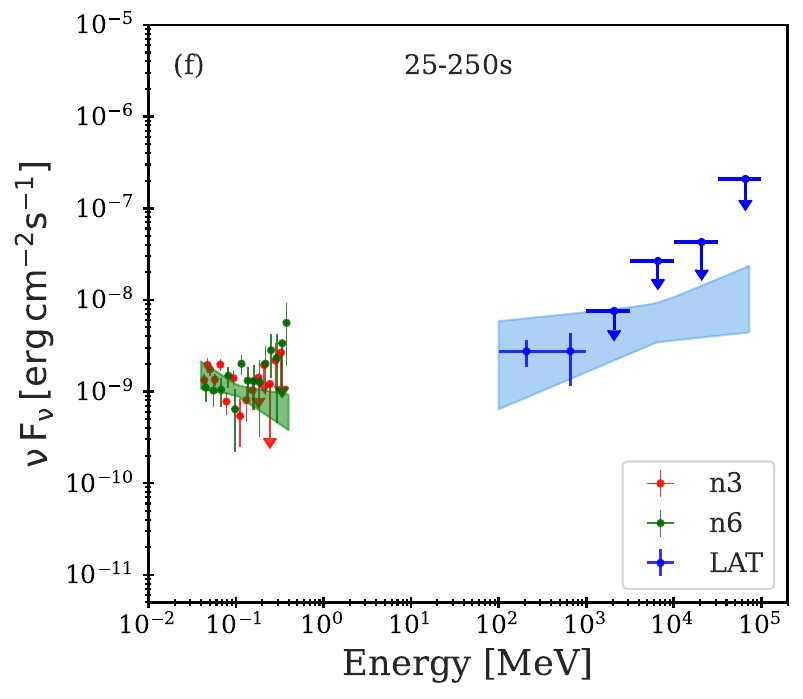} 
\caption{Temporal evolution of broadband SED of GRB~230812B from \textit{Fermi}/GBM and \textit{Fermi}/LAT analysis. The green band depicts the best fit model in keV-MeV band resultant from \textit{Fermi}/GBM analysis (see table \ref{tab:GBM_results}) extended up to 100 GeV. The blue butterflies shows the powerlaw model fit on \textit{Fermi}/LAT spectra in 0.1-100 GeV (see table \ref{tab:LAT}). Forward-folded flux data points from the NaI and BGO detectors, as well as from LAT, are shown as filled circles. Panels (a)–(c) correspond to the prompt emission phase (Bins 1-3). Panel (d) represents the prompt-to-afterglow transition (Bin 4) Panels (e)-(f) show the afterglow phase (Bins 5–6). The overall SED evolution shows the emergence and evolution of the spectral components}

\label{fig:prompt_sed}
\end{figure*}

\subsection{Radio} We analysed the uGMRT data obtained through our Director’s Discretionary Time (DDT) proposal (ddtC304; PI: Shraddha Mohnani). Three observations, each with a 2-hour time slot, were conducted on 17 September 2023 and 30 September 2023 using Band-5 (1050–1450 MHz), and on 1 October 2023 using Band-4 (550–850 MHz).

We processed the interferometric data using CAsa Pipeline-cum-Toolkit for Upgraded Giant Metrewave Radio Telescope data REduction \citep[CAPTURE\footnote{\url{https://github.com/ruta-k/CAPTURE-CASA6}};][]{CAPTURE}. CAPTURE follows the standard procedure for interferometric data reduction, which includes initial flagging of bad data, followed by delay, bandpass, flux, and complex gain calibration. The flux density scale was set using the Perley–Butler 2017 scale.
After the initial calibration stage, we manually inspected the data and flagged any remaining corrupted data using rFlag and TFcrop. We then performed an additional round of calibration. The derived calibration solutions were applied to the target source, and the data were subsequently split for self-calibration. Three rounds of phase-only self-calibration were carried out, using gain solutions derived from the target itself to correct for phase errors. Final imaging was performed using WSClean \citep{wsclean_2014}. For the 17 September Band-5 observation, we achieved an rms noise level of approximately $\rm{\sim14\,\mu Jy/beam}$. However, no significant emission was detected at the location of the target source. We reported our analysis results in NASA’s General Coordinates Network (GCN) circular \citep{GCN34727}. To improve sensitivity, we combined the Band-5 datasets from the observations conducted on 17 and 30 September 2023. However, no radio counterpart was detected in the combined image either. At band 4, we reached an RMS noise of (write the value). However, no detection was noted in band 4 as well.

\section{Results}\label{results}

\subsection{Recovering the afterglow from keV to GeV}
 Given that GRB~230812B is exceptionally bright (with fluence of $2.52\rm{\times 10^{-4} \,erg/cm^2}$), we modeled the \textit{Fermi}/GBM background using OSV (see Sect.~\ref{sec:gbm}). We estimated the flux and the spectral parameters of the faint MeV afterglow emission in the energy band of 40-400\,keV up to 650\,s post burst. The spectra
beyond T$_{90}$ required PL spectral model except for Bin-4 (3.6-10.0\,s), where the spectrum preferred a BPL. The spectral peak for Bin-4 is estimated to be $99.2\pm3.1$\,keV. For Bin-5, -6, and -7, the spectra is fitted with a PL with indices around -2.3 providing a bound on the peak energy less than 40\,keV. The spectra for different bins are presented in Figure~\ref{fig:prompt_sed} and the corresponding spectral parameters are presented in Table~\ref{tab:GBM_results}.

The joint observation performed by \textit{Fermi}/GBM and \textit{Fermi}/LAT allows to cover a wide energy range from 8\,keV up to 100\,GeV (see Figure \ref{fig:prompt_sed}) until 1\,ks from the onset of the burst. During the prompt emission (until the end of T$_{90}$; 3.6\,s), the spectra observed with \textit{Fermi}/GBM are best described by the Band function \citep{Band_function}, whereas the \textit{Fermi}/LAT spectrum is fitted with a power-law model. The \textit{Fermi}/LAT spectrum during the prompt emission is softer than -2 and can be identified as a tail of the \textit{Fermi}/GBM spectrum extrapolated to GeV energies (see Figure~\ref{fig:prompt_sed}). In Bin-3 and -4, the spectra in \textit{Fermi}/LAT can be identified as an additional spectral component given the softer $\beta$ (see Table~\ref{tab:GBM_results}) hinting towards the emergence of a second component. 
Bin-4 is better described with BPL\footnote{We tested the following models in search for a the best fit model: PL, SBPL, Band, Band+PL, and BPL.}.

We find that for the afterglow bins, the \textit{Fermi}/GBM spectrum is best described with PL indices softer than 2.2 (in the range 2.2-2.5; see Table~\ref{tab:GBM_results}). However, \textit{Fermi}/LAT shows a spectral hardening (Bin-6) with a photon index of $1.75\pm0.15$. 

\begin{figure*}[ht!]
    \includegraphics[width=\textwidth,]{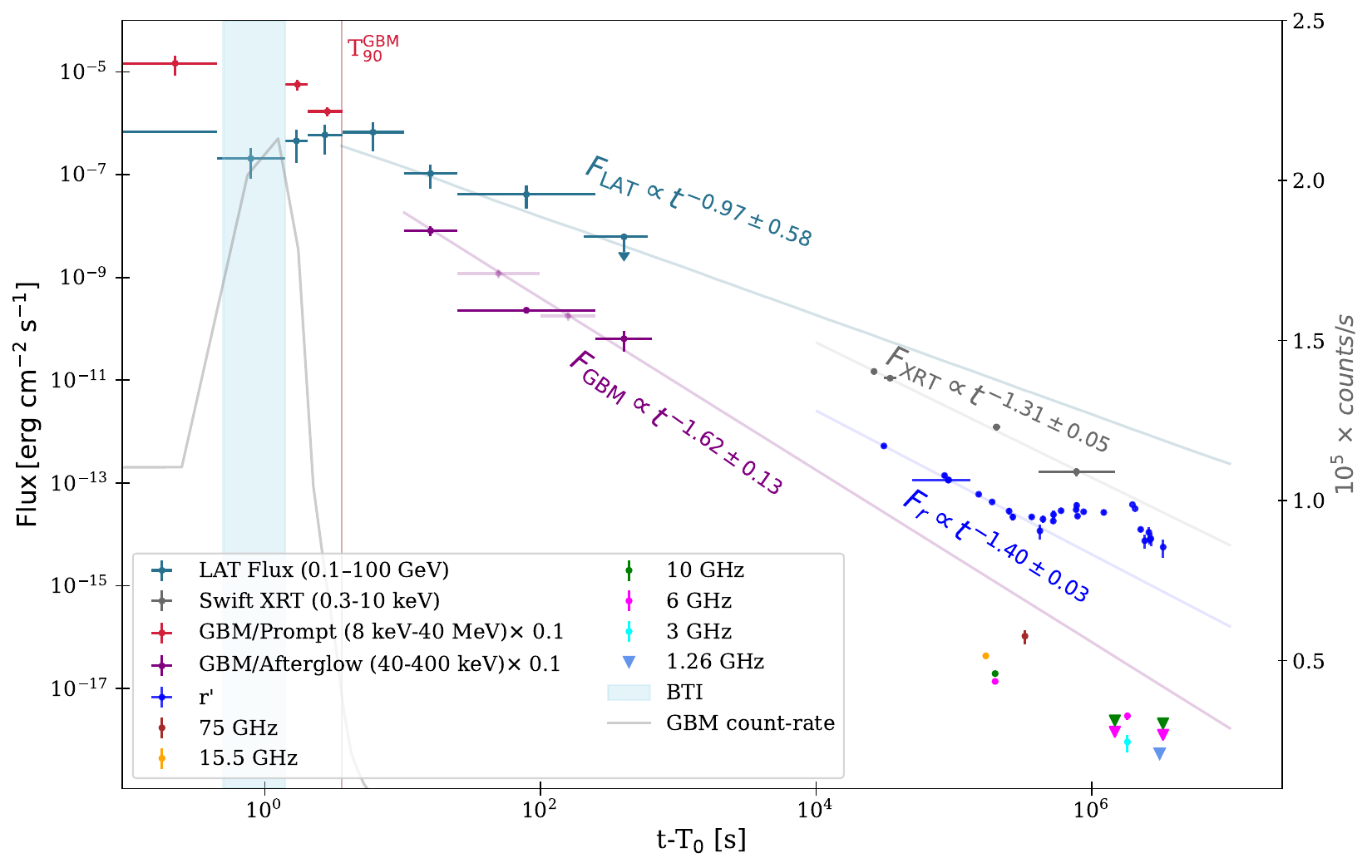}
\caption{Multiwavelength flux lightcurve (left y-axis) of GRB~230812B obtained from \textit{Fermi}/LAT (0.1-100 GeV), \textit{Fermi}/GBM prompt (8 keV-40 MeV) and afterglow (40-400 keV), \textit{Swift}/XRT (0.3-10 keV), optical (r$^{\prime}$) and radio (at various frequencies ranging from 1.2 to 75 GHz) telescopes. The GBM afterglow data is rescaled by factor of 0.1 for plotting purpose. The faint data points correspond to fluxes obtained from finer time binning. The r$^{\prime}$ band data shows the optical afterglow and the emission from the associated supernova. The solid lines corresponds to the best fit power-law model F=F$_{0} (t-{T_0})^{-\Gamma_{t}}$ for afterglow data from each instrument. For each instrument the time interval used for the fit along with the powerlaw index obtained as result are given in Table \ref{tab:LC_fit}. The grey solid curve shows the count-rate light curve with rates depicted by the right y-axis.} 

\label{fig:mwlc}
\end{figure*}
\subsection{Temporal Evolution: LC fitting}
\label{sec:LC_model}

To infer the temporal evolution of GRB~230812B, we fitted the light curves from \textit{Fermi}/GBM, \textit{Fermi}/LAT, \textit{Swift}/XRT and optical (r$^{\prime}$-band) with a power-law model $\rm{F = F_{0}\,(t-{T_0})^{-\Gamma_{t}}}$ as shown in Figure~\ref{fig:mwlc}. We used Bayesian inference to estimate the temporal decay index ($\Gamma_{t}$) and the normalization factor (F$_0$). To estimate the posterior distributions of these parameters, we used a Markov Chain Monte Carlo (MCMC) using the emcee sampler \citep{emcee}. The likelihood function was defined by assuming Gaussian errors in the flux measurements $\rm{\ln \mathcal{L} = -\frac{1}{2} \sum_{i} \left[ \frac{(F_i - F_{\mathrm{model},i})^2}{\sigma_i^2} + \ln(\sigma_i^2) \right]}$ and the uniform priors for the parameters were used. 

Bin-4 has been avoided from the GBM and LAT light curves while performing the temporal fit due to its close proximity to the prompt emission phase. 
For the optical light curve, we restricted the fit to data prior to the emergence of the associated supernova (31-251\,ks).

We found that the indices of temporal decline are similar for the multiwavelength light curves (in optical, X-rays and GeV energies) from $\sim$10\,s onward extending up to about $\sim 10^{6}$\,s. 
The individual temporal decay indices of the afterglow light curves resulting from the fit are reported in Table~\ref{tab:LC_fit}. Notably, for LAT, XRT and r$^{\prime}$, the temporal decay indices are consistent within statistical uncertainties. However, the early afterglow (within 10\,ks) observed with \textit{Fermi}/GBM in 40-400\,keV is slightly steeper. Moreover, the data do not suggest the presence of the jet-break in optical, X-ray and GeV energies.

\begin{table*}[ht]
    \centering
    \begin{tabular}{cccccccr} \hline
Bin &t - T$_{0}$ & E$_{\rm max}$    &\multicolumn{2}{c}{Flux }            &\multicolumn{2}{c}{Index }     & TS  \\ 
 &$\rm [s]$ &[GeV] & \multicolumn{2}{c}{[10$^{-7}$erg cm$^{-2}$ s$^{-1}$]}& & &  \\ 
 \cmidrule(lr){4-5} \cmidrule(lr){6-7}
   & &  &0.1 - $\rm{E_{max}}$ &  0.1 - 100\,GeV                      &0.1 - $\rm{E_{max}}$  & 0.1 - 100\,GeV& \\ \hline
 
 Bin-1 & 0 - 0.4       & 0.2   &  $6.73 \pm 2.82$  & $6.38 \pm 2.43$  & 4.01 $\pm$ 1.07   &  --  &            63  \\
 Bin-2 & 1.4 - 2.0      & 0.3 &  $4.54 \pm 2.82$  & $3.76 \pm 1.74$  & 2.83 $\pm$ 0.75 &  -- &            41  \\
 Bin-3 & 2.0 - 3.6   & 0.9  &  $5.94 \pm 3.51$  & $4.37 \pm 1.50$  & 2.37  $\pm$   0.39  & $1.96 \pm 0.52$   & 121   \\
 Bin-4 & 3.6 - 10.2  & 2.5  &  $6.70 \pm 3.86$  & $3.38 \pm 0.10$  & 1.90   $\pm$ 	0.20 & $1.62 \pm 0.26$& 202  \\ 
 Bin-5 & 10.2 - 25.0   & 1.1 &  $1.05\pm0.52 $   &   $0.76\pm0.22$    & 2.33  $\pm$    0.31      &$1.97\pm0.40$ & 94  \\
 Bin-6 & 25.0 - 250.0  &  72.0 &  $0.41\pm0.19 $    &    $0.38\pm0.17$    & 1.75  $\pm$ 0.15    &$1.75\pm0.15$    & 136 \\
 Bin-7 & 251.9 - 647.2 & 14.0 & $<$ 0.51   & $<$ 0.16                    & 1.47  $\pm$ 0.52     & $0.01 \pm 0.003$ & 19  \\
\hline
    \end{tabular}
    \caption{Results of the \textit{Fermi}/LAT unbinned likelihood analysis for GRB~230812B using a power-law model. The energy flux and spectral index are reported separately for two energy ranges: 0.1–100 GeV, and 0.1 GeV up to the highest-energy photon ($\rm{E_{max}}$) detected in each time bin (as listed in the second column). Flux values corresponding to a TS below 25 are reported as upper limits ($\rm{3\sigma}$).}
    \label{tab:LAT}. 
\end{table*}

\begin{table}[ht]
    \centering
    \begin{tabular}{ ccc} \hline
t-T$_{0}$    &Flux (0.3-10\,keV) &  Index                    \\ 
 $\rm [ks]$       &  [10$^{-14}$ erg cm$^{-2}$ s$^{-1}$] &   \\ \hline
 25.4-27.2 &  $ 1479 \pm 102 $  &  $1.82\pm0.10$                \\
 30.9-38.2 & $ 1096 \pm 76 $  & $1.65\pm0.10$                \\ 
 191.7-215.6 &  $ 123\pm 20 $ &  $1.84\pm0.22$               \\
 407.8-1466.0 & $ 17\pm3 $  & $1.72\pm0.24$                   \\
\hline
    \end{tabular}
    \caption{Time resolved spectral analysis results of x-ray afterglow of GRB~230812B from \textit{Swift}/XRT in 0.3-10 keV band. Data for all four time-bins is simultaneously fit with powerlaw model taking into account the galactic and intrinsic metal absorption (N$_{\rm H}$(z) = 10$^{22}$ cm$^{-2}$).}
    \label{tab:Swift-XRT}
\end{table}

\begin{table}[ht]
    \centering
    \begin{tabular}{ cccc} \hline
Instrument & Energy Band  & t-T$_0$   & $\rm{\Gamma_{t}}$ \\  \hline
\textit{Fermi}/LAT& 0.1-100 GeV  & 10.2-647 s & $-0.97\pm0.58$   \\
 \textit{Fermi}/GBM& 40-400 keV & 10.2-647 s & $-1.62\pm 0.13$  \\ 
\textit{Swift}/XRT & 0.3-10 keV  & 25-1400 ks  &  $-1.31\pm0.05$ \\
Optical & r$^{\prime}$  & 31-251 ks &  $-1.40\pm0.03$\\ \hline
    \end{tabular}
    \caption{Light curve fitting results of GRB~230812B from various instruments (column 1), in their respective energy bands (column 2). The fits are performed over the time intervals specified in the third column. A power-law model of the form  F=F$_{0} (t-{T_0})^{-\Gamma_{t}}$, is used, with the best-fit temporal indices $\Gamma_{t}$ reported in the last column. The corresponding best-fit models for each instrument are shown in Figure~\ref{fig:mwlc}.}
    \label{tab:LC_fit}
\end{table}

\begin{figure*}[ht!]
    \includegraphics[width=\textwidth,]{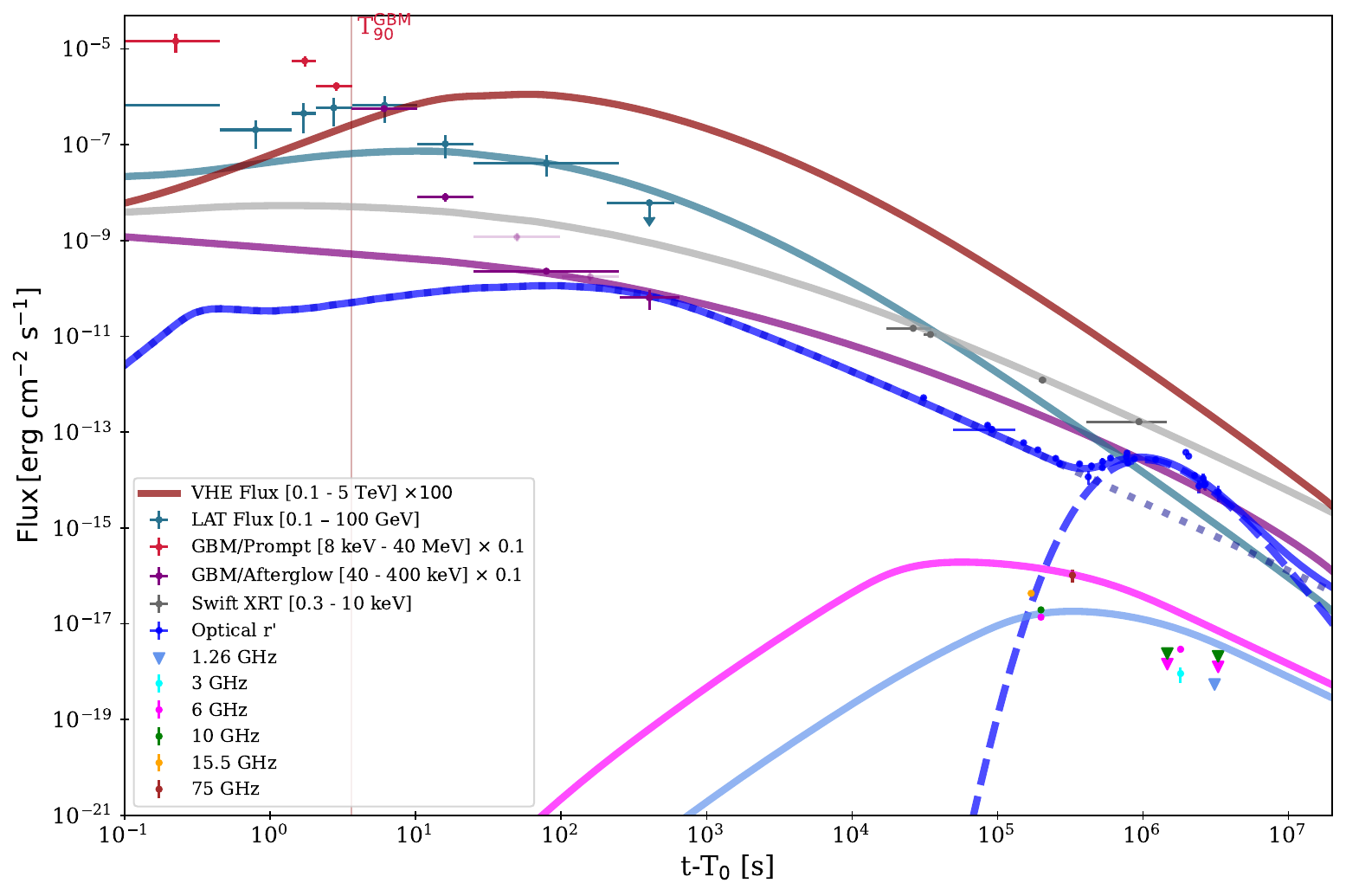}

\caption{Modeling of the multiwavelength light curves with synchrotron and SSC scenario compared with early and late time afterglow observations for GRB~230812B. Predictions for the model are displayed in solid colored lines. The corresponding instruments and energy bands are listed in the legend of the plot (see Sect.~\ref{sec:MWLdata}).}
\label{fig:lc_pred}
\end{figure*}

\begin{figure*}[ht!]
    \includegraphics[width=\textwidth,]{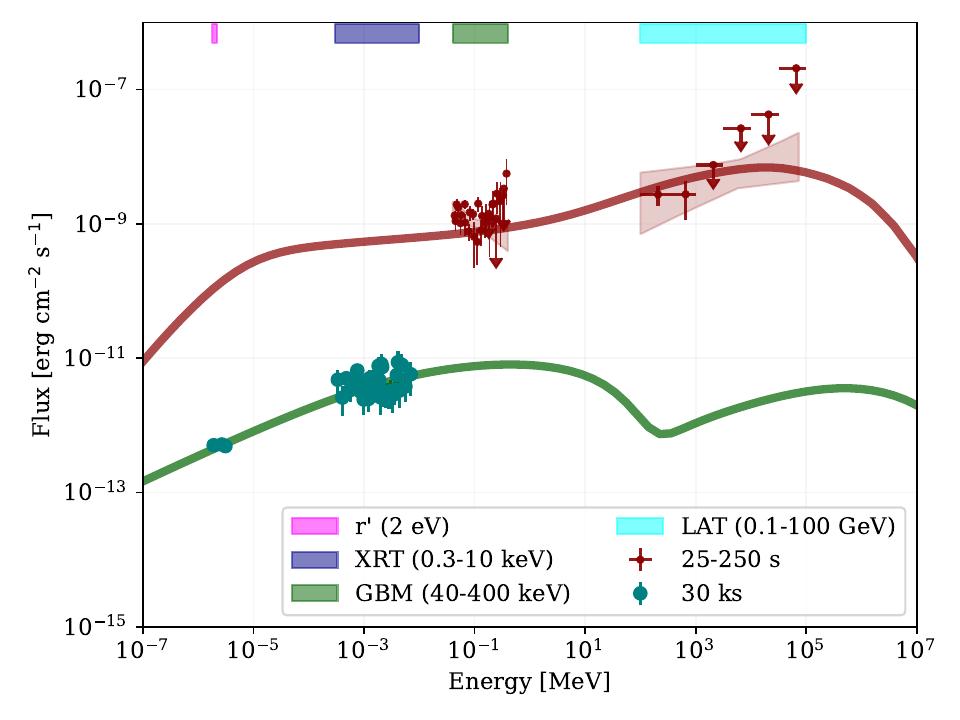}
\caption{Modeling of the GRB~230812B broadband spectral energy distribution (solid red and green lines) for two different time intervals including early-time X-ray and high-energy afterglow data (25-250 s) and late-time afterglow optical and X-ray data at 30 ks. In the modeling afterglow emission is produced in the synchrotron and SSC external forward shock scenario. Observed data points (filled circles) and upper limits (arrows) from multiple instruments spanning the early to late afterglow phases are displayed. The energy bands used in the analysis are indicated in the legend: optical (r$^{\prime}$; magenta), XRT (0.3-10 keV; purple), GBM (40-400 keV; green), and LAT (0.1-100 GeV; cyan). Details of the modeling and derived parameters are provided in Sect.~\ref{sec:spec_model}.}

\label{fig:ssc_model}
\end{figure*}

\section{Theoretical modeling and interpretation}\label{sec:spec_model}

 A more general interpretation of the broadband emission of GRB~230812B can be obtained by combining together the whole observational data as shown in the SEDs and in the light curves in Figure \ref{fig:prompt_sed} and Figure \ref{fig:mwlc}. We interpret this broadband emission in the context of the standard external forward shock afterglow scenario. In this model, the two radiation components are produced by synchrotron and SSC mechanisms and they can explain the multi-wavelength emission \citep{Sari_Piran_Narayan1997,Sari_Esin_2001,Nakar2009}. The model used for the interpretation was presented in \cite{Miceli2022}. The jet dynamics follows the approach proposed in \cite{Nava_2013} in the homogeneous shell approximation. As a result, the only free parameters assumed are the initial Bulk Lorentz factor $\Gamma_0$ and the afterglow kinetic energy $E_{k}$. For the circumburst environment, we considered a wind-like scenario with density $n = 3\times 10^{35} A_\star R^{-s}$ with $s = 2$ and $A_\star$ being the density normalization for wind-like environments \citep{Chevalier_2000,Panaitescu_2000}. We assumed that constant fractions of the kinetic energy of the blastwave are transferred to electrons ($\epsilon_e$) and to amplify the magnetic field ($\epsilon_B$). The electrons swept up by the shocks are assumed to be accelerated into a power-law distribution $dN / d\gamma \propto \gamma^{-p} $ where $p$ is the electron spectral index and $\gamma$ is the electron Lorentz factor. 
 
\subsection{Modeling of multi-wavelength light curves}

\label{subsec:mwl_model_lc}
 The resulting predicted afterglow emission covering the whole multiwavelength range from radio to VHE and the time intervals from $0.1$ s up to $10^7$ s is shown in Figure \ref{fig:lc_pred}. The values of the afterglow free parameters that best reproduce the observations are $E_k = 8 \times 10^{52}$ erg, $\Gamma_0 = 70$, $A_\star = 0.2 $, $\epsilon_e = 0.1$, $\epsilon_B = 5 \times 10^{-5}$ and $p = 2.2$. The model suggests that afterglow emission starts to dominate in the GBM band from around $25$ s, while LAT data seem to be consistent with afterglow even at earlier time, from $\sim 5-10$ s, as also seen from the time-binned spectra in Figure \ref{fig:prompt_sed}. The modeling is able to consistently reproduce both the \textit{Fermi}/LAT and \textit{Fermi}/GBM observational data collected starting from $\sim 25$ s up to $t < 10^3$ s and the X-ray and optical data collected from $\sim 10^4$ s up to $\sim 10^6$ s. No evidences for jet breaks or other dominant component are present until t $\sim 10^6$ s. For $t \gtrsim 3 \times 10^5$ s the optical data are dominated by the emission due to the supernova. We therefore added this contribution to the predicted optical afterglow. Radio data have been collected at late times from $ \sim 10^5$ s up to $ \sim 3 \times 10^6$ s for different frequencies, from 1.26 GHz up to 75 GHz. We produced the predicted light curves for two reference frequencies, i.e. 1.26 GHz and 6 GHz. In this case, the model predictions overestimate the observational data points and upper limits by a factor $\sim 10$ even though the expected decay of radio flux in an optically thin regime from $\sim 10^5$ s up to $\sim 10^6$ s is reproduced. Multi-wavelength data can provide important clues on the free parameter of the GRB afterglow model and on the break frequencies (self-absorption frequency $\nu_{sa}$, minimum frequency $\nu_m$ and cooling frequency $\nu_c$) of the synchrotron spectrum. Assuming that the emission is produced in slow cooling ($\nu_m$ < $\nu_c$), the adopted scenario for late-time data requires to have a hard value of $p$ which is adopted here to be $2.2$ to reproduce the X-ray and optical time-evolution following also the analytical prescription of \cite{2002ApJ...568..820G}. The radio detections at $10^5$ s followed by the upper limits or marginal detections at $10^6$ s can be interpreted as the emission produced in an optically thin regime ($\nu_{sa} < \nu_{radio}$) with the minimum frequency $\nu_m$ crossing the radio band during this time interval. On the other hand, the \textit{Fermi}/GBM and \textit{Fermi}/LAT early-time data can constrain the free parameters of the GRB dynamics. In particular, the afterglow kinetic energy $E_k$ cannot be larger than $8 \times 10^{52}$ erg, implying a prompt efficiency $\eta \gtrsim 0.55$. Assuming the deceleration peak to be around 5-10 s when the afterglow component seems to start rising in \textit{Fermi}/LAT spectra and light curves, we can derived acceptable values for $\Gamma_0$ in the range between 70-100. Concerning the microphysical parameters, in the modeling we adopted $\varepsilon_e = 0.1$ and $\varepsilon_B = 5 \times 10^{-5}$ which are consistent with similar results obtained in modeling of GRBs detected in the HE and VHE domain \citep{Gao2015,epse_epsb_vhe}.

\subsection{SED modeling}

In Fig.~\ref{fig:ssc_model} we compare the modeling discussed in the previous sections with the SED data at two different times. To cover the time evolution of the source, we consider two time intervals representative respectively of the early and late time emission (see Figure \ref{fig:ssc_model}): (i) the combined spectrum observed by \textit{Fermi}/GBM (40--400\,keV) and \textit{Fermi}/LAT (0.1--72\,GeV) in the $25 - 250$ s time interval; (ii) the late-time afterglow spectrum including optical and X-ray data at 30 ks.

According to the  model, the early-time \textit{Fermi}/LAT emission is dominated by the SSC component of the spectra peaking in the GeV energy range and the \textit{Fermi}/GBM emission is dominated by synchrotron emission in the limit just below the high-energy cutoff of the synchrotron spectrum and the rising of the SSC component. The minimum frequency $\nu_m$ is located in the optical band at few eV, as also confirmed by the change of slopes in the predicted optical light curve at $\sim 250$ s as seen in Figure \ref{fig:lc_pred}.

The late-time optical and X-ray data at 30 ks are instead consistent with the synchrotron spectrum in the slow-cooling scenario for frequencies $\nu_{m} < \nu < \nu_c$ assuming a value of $p = 2.2$ which implies a $F_{\nu} \propto \nu^{-0.6}$. The cooling frequency $\nu_c$ is located above the X-ray data at around $\sim 50-100$ keV and the minimum frequency $\nu_m$ is at $\sim 0.5$ THz (not visible in the plot). This interpretation is also in agreement with the optical and X-ray time evolution observed from the light curve in Figure \ref{fig:lc_pred}. The evolution of the modelled spectra at different times shows that the peak moves towards higher frequencies with time, as expected in the wind scenario. In conclusion, both time intervals are consistently explained with the synchrotron and SSC afterglow external forward shock scenario.

\begin{figure*}[ht]
    \centering
    \includegraphics[width=0.49\textwidth,]{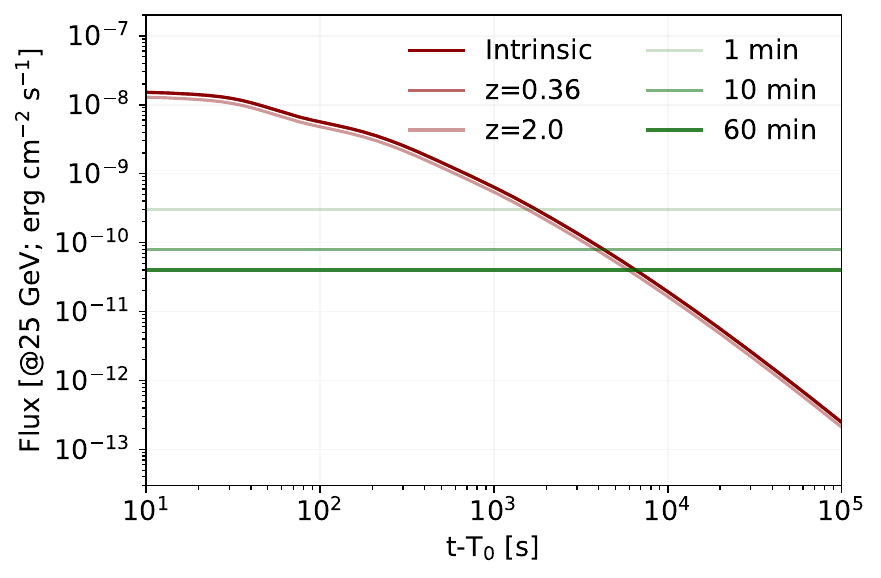}
    \includegraphics[width=0.49\textwidth,]{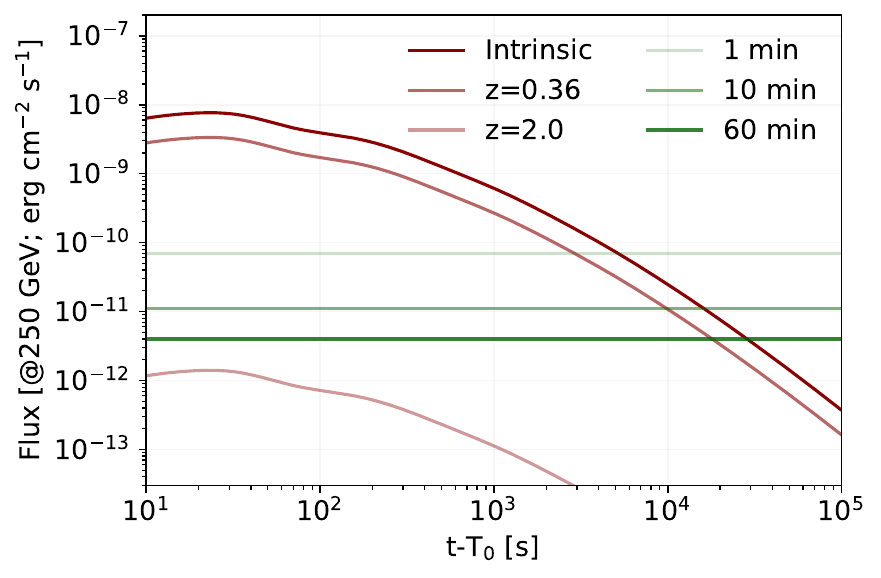}
    \caption{The CTAO-North array detectability of the predicted gamma-ray emission at 25 GeV (left plot) and 250 GeV (right plot) for GRB~230812B (z=0.36) and for a GRB~230812B-like event but with a larger redshift of  z=2. The red lines represent the intrinsic GRB emission (dark-red), EBL-corrected observed emission after EBL at redshift z=0.36 (red) and at redshift z=2.0 (light-red). The horizontal lines are the differential sensitivity of CTAO with different exposures: 1\,min: light-green, 10\,min: green, and 60\,min: dark-green.}
    \label{fig:EBL}
\end{figure*}

\subsection{Predicting VHE afterglow and detectability}

The observation of a 72 GeV photon and the modeling of the LAT emission component in the 0.1-100 GeV energy range provide compelling evidence for the presence of a SSC component in GRB~230812B. This result make this event particularly interesting to test a possible detection in the tens-hundreds GeV energy band by IACTs. Unfortunately, none of the active ground-based instruments have reported the detection of the event. Nevertheless, rather than the intrinsic absence of a VHE component, this could be easily explained considering the low duty cycle of IACTs and the large localization error of the event in the first minutes that prevent to perform a rapid follow-up. We provide an estimate of the expected intrinsic flux for two energies values, 25 GeV and 250 GeV, respectively representative of the lowest energy threshold for IACTs and of the typical energy value in standard VHE observations. We extract the intrinsic flux at these energies and its time evolution using the same model inputs provided in Sect.~\ref{sec:spec_model}. Then, we correct the intrinsic fluxes for the attenuation due to the extragalactic background light (EBL) using the EBL model of \cite{Dominguez} to estimate the expected observed energy flux. We assume two values of the redshift: the one estimated for GRB~230812B (z = 0.36) and a more distant redshift (z = 2) assuming the same intrinsic properties of GRB~230812B. The intrinsic and observed light curves for the two energy values are shown in Figure~\ref{fig:EBL} with different shades of red. We compared the expected VHE fluxes with the Cherenkov Telescope Array Observatory northern array differential sensitivities\footnote{Extracted from \url{https://www.cta-observatory.org/science/ctao-performance/\#1472563157332-1ef9e83d-426c}}(CTAO-North) assuming three different exposure times ($1$ min, $10$ min and $60$ min, in Figure ~\ref{fig:EBL} marked in different shades of green) to explore the detectability of GRB\,230812B by IACTs in the HE and VHE band. 

For the 25 GeV energy value, the flux in the early afterglow (within 1\,ks) is above the sensitivity limit even considering a very low exposure of 1\,min for the CTAO-North array and it extends up to $\sim 5$ ks for an exposure of 10-60 min. In addition, the limited impact of the EBL at this energy is evident especially when considering the observed flux at redshift z = 2. As a result, larger distances did not significantly impact on the detection capabilities. For the 250 GeV energy value, results are slightly better for z = 0.36. The predicted flux is above the CTAO-North sensitivities for a 1\, min exposure up to $\sim 1$ ks and for 10 or 60 min exposures up to $\sim 10$ ks. On the other hand, at redshift z = 2, the strong EBL absorption largely suppresses the flux which results to be more than one order of magnitude fainter than the diffrential sensitivities. These results, especially the ones obtained for the 1\, min exposures, confirm that having the source in the FoV of CTAO-North array, even for a very limited time interval which is compatible with a tiling strategy, could have resulted in a confirmed detection. 
However, relying on precise localization is problematic. Indeed, the arc-second localization of the burst, provided by \textit{Swift}/XRT, was only available 30~ks after triggering, too late for effective GRB detection.

In order to understand the possibility of following up the GRB with IACTs, we explored the Healpix localization maps\footnote{The Healpix maps contain the final localization information of the GRB which is distributed as a final notice by \textit{Fermi}/GBM of a burst (\citealt{2023GCN.34386....1F}; \url{https://heasarc.gsfc.nasa.gov/FTP/fermi/data/gbm/triggers/2023/bn230812790/quicklook/}glg\_healpix\_all\_bn230812790.fit).} through TilePy software \citep{Seglar-Arroyo:2024nap}, an open-source Python package designed for the automatic scheduling of follow-up observations of transient events. Our estimates show that the GRB was visible to CTAO/LST for about one hour from the location of CTAO-North. In this following section, we motivate that even if the precise localization was not known, a systematic tiling strategy might have resulted in a detection in VHE $\gamma$-rays.

We note that the first announcement of GRB~230812B (trigger number: 713559497\footnote{\url{https://gcn.gsfc.nasa.gov/other/713559497.fermi}}) was communicated through GCN notices at Sat 12 Aug 2023 18:58:19 UT (T$_{0}$; notice-1). The trigger was later updated by around 20\,s, confirming that it originated from a GRB (T$_{1}$: Sat 12 Aug 2023 18:58:42 UT) with probability of 96\% (notice-2). In notice-2, within $\sim$20\,s, the co-ordinate of the GRB was communicated with R.A. 248.867$^\circ$ and Dec. +41.917$^\circ$ (L$_{1}$) and with associated error-radius of 3.22$^{\circ}$ (statistical only). In a follow-up notice (notice-3) within about 120\,s (T$_{2}$: Sat 12 Aug 2023 19:00:12 UT), a more precise localization has been distributed with a revised R.A. 249.720$^\circ$ and Dec. +45.970$^\circ$ (L$_{2}$) with an error-radius of 1$^{\circ}$ (statistical only). This location is $\sim$1.9$^{\circ}$ away from the final localization of the GRB as estimated by the XRT observations. 

Since, the GRB was visible from the CTAO-North site, we propose a rapid follow-up strategy with CTAO/LST of the Fermi GBM localization. We present this GRB as a hypothetical test case. Depending on the visibility in the sky (if the source is visible from the location of the CTAO) we propose to slew the telescope to the center of the error-region (in this case to L$_{1}$) once the trigger is identified coming from a GRB (with probability of being a GRB to be more than 50\%, usually announced in notice-2). The communication of the trigger and the slew time can reasonably be approximated to be around 60\,s (circulation of the second notice: notice-2 and the reaction time of the telescope). This number includes around 30 s for the trigger communication and 30 s for the slew time. We propose to perform a minute long observation centering at L$_{1}$ and initiate a tiling observation, as also generally proposed in search for electromagnetic counterpart of gravitational waves. The tiling should continue until the next notice (usually with a more precise localization) arrives. In our case, the next notice (notice-3) arrives around 120\,s from the trigger time. Hence, for this case, after a minute long observation at L$_{1}$, the telescope needs to be moved to L$_{2}$ which might introduce a delay due to the slew rate of LST\footnote{Usually the slew rate is around 120$^\circ$ per 20\,s. This rate implies that a slew of about 3$^{\circ}$ should take about a fraction of seconds, which can be neglected.}. Once the telescope reaches the location L$_{2}$, we propose to make a tiling of the 3$\sigma$ error region, which in this case is 3$^{\circ}$ (1$\sigma$ error-region is 1$^{\circ}$). The corresponding sky-localization is about 28.3 deg$^{2}$. Given the FoV of LST of 5\,deg$^{2}$, a total of about 6 points should be made in order to cover the sky-localization patch. If 1 min is spent as an exposure (t$_{\rm exp}$) for each tiling observation, the total time that is required to cover the 3$\sigma$ error-region on notice-3 is about 360\,s. 
A total latency of performing the observations proposed above is 510\,s from the trigger time.
We note from Figure~\ref{fig:EBL} left (right) panel that the source at 25 GeV (250 GeV) is brighter than the limiting flux for a detection even beyond 510\,s and could have been discovered with LST. Moreover, the source, if placed at higher redshift, might have also been detected at lower energies (such as 25 GeV) because the spectrum is almost unaltered (left panel of Figure~\ref{fig:EBL}) by EBL absorption as opposed to 250 GeV (right panel of Figure~\ref{fig:EBL}).

\section{Discussion and conclusion} \label{discussion} 
GRB~230812B is one of the brightest $\gamma$-ray bursts observed by \textit{Fermi}/GBM, with a high fluence of $3.27\rm{\,\times\, 10^{-4} \,erg/cm^2}$ \citep{GCN34391}. Due to the large flux, the \textit{Fermi}/GBM detectors experienced a pulse pileup for $\sim1$s. GRB~230812B was detected by \textit{Fermi}/LAT from the onset of the prompt emission and continued up to 1\,ks extending well into the afterglow phase. \textit{Fermi}/LAT recorded the highest-energy photon of 72\,GeV at around 32\,s after the trigger. \textit{Swift}/XRT recorded the X-ray afterglow of the burst $\sim25$ ks post trigger \citep{GCN34400}. Subsequently, various optical telescopes monitored the burst, successfully detecting the optical counterpart and the associated supernova SN 2023pel \citep{GCN34597}. In the radio band, confirmed detection in the early days (< 17 days) and flux upper limits later, have been obtained by observations made by AMI-LA \citep{GCN34433},  VLA \citep{GCN34552, GCN34735}, and uGMRT \citep{GCN34727} telescopes.

By collecting background photons from Fermi orbits having the same geographical footprints, preceding and following the observation of the $\rm{\gamma}$-ray burst, we estimated the average background spectrum using the orbital subtraction tool \citep{osv_tool}. We used custom response matrices generated for each time interval and were able to recover hard X-ray emission beyond T$_{90}$. The hard X-ray spectra after 3.6\,s were best described by a powerlaw model. The \textit{Fermi}/GBM afterglow was extracted in the 40-400\,keV energy band, which corresponds to the most sensitive range of the NaI detectors. In the absence of any detection with \textit{Swift}/BAT, recovering the hard X-ray spectra following this method is pivotal because of the possibility of increasing the number of early-afterglow detections in GBM-detected GRBs. 

The temporal index $\Gamma_t$ of light curve during the afterglow (assuming F $\rm{\propto(t-{T_0})^{-\Gamma_{t}}}$ seen in the energy band of 40-400\,keV is  $1.62\pm0.12$. The steeper temporal index might be a consequence of the inclusion of the temporal bin 10-25\,s which is close to the prompt emission phase and the prompt-contamination in this time bin can not be ignored. We identify a similar temporal decline in X-rays (0.3-10\,keV), optical (r$^{\prime}$ band) and high-energy (HE) $\gamma$-rays.

The combined \textit{Fermi}/GBM and \textit{Fermi}/LAT early-time (t $< 10^3$ s) data together with the late-time (t $> 10^3$ s) X-ray, optical and radio data allows us to build broadband light curves and SEDs and hence to model the afterglow emission and its time-evolution in a wide range of frequencies.  We used the numerical modeling presented in \cite{Miceli2022} to derive the predicted time-evolving broadband lightcurves of GRB230812B  from $0.1$ s up to $10^7$ s (see Figure~\ref{fig:lc_pred}) and the SEDs for two selected time intervals ($25 - 250$ s; $30$ ks, see Figure \ref{fig:ssc_model}) in the afterglow external forward shock scenario. We also add the prediction of the expected light curve in the VHE domain (0.1 - 5 TeV) for a typical energy range of observation. For the circumburst environment, we adopted a wind-like scenario considering that the GRB progenitor is a collapsing massive star, as confirmed by the associated supernova detection \citep{GCN34597, SN_paper_GRB230812B}. As a result, the choice of the wind-like environment, rather than a constant-density environment, is justified \citep{2000ApJ...536..195C}. The values of the afterglow free parameters that best reproduce the observations are $E_k = 8 \times 10^{52}$ erg, 
 $\Gamma_0 = 70$, $A_\star = 0.2 $, $\epsilon_e = 0.1$, $\epsilon_B = 5 \times 10^{-5}$ and $p = 2.2$. 

Data collected from optical up to HE are consistently reproduced with the afterglow external forward shock scenario without any evidence of jet breaks or other dominant components until $\sim 10^6$ s, with the only exception of the emission produced by the supernova which dominates optical data starting from $\sim 3 \times 10^5$ s.  Our modeling suggests that afterglow emission starts to dominate the GBM and LAT data respectively at $\sim 25$ s and $5-10$ s. In addition, GBM and LAT observations can be exploited to constrain the free parameters of the GRB dynamics, in particular the afterglow kinetic energy E$_k$ and the bulk Lorentz factor $\Gamma_0$. The time-evolution of the X-ray and optical data provide evidences for a value of $p = 2.2$. The other microphysical parameters ($\varepsilon_e$ and $\varepsilon_B$) are consistent with similar modelings of TeV-detected GRBs. From the SED modeling, we demonstrate that the sub-MeV afterglow emission detected by GBM arises from synchrotron radiation emitted just below the expected high-energy cut-off, while the LAT emission in the GeV domain is purely associated with the SSC component peaking in the GeV domain in the $ 25 - 250$ s time interval. The late-time optical and X-ray data are consistently reproduced assuming the synchrotron radiation in the slow cooling scenario ($\nu_m < \nu < \nu_c$).

The detection of a 72 GeV photon presents compelling evidence for the existence of a VHE emission component, as observed in a few GRBs up to now. We estimated the expected intrinsic and EBL-corrected observed VHE light curves at 25 GeV and 250 GeV for GRB\,230812B (z=0.36) and for a simulated event with the same intrinsic properties of GRB\,230812B but assuming a larger redshift ($z = 2$). The choice of $z = 2$ is justified as the redshift distribution of long-GRBs peaks at around $z = 2$ \citep{Ghirlanda:2022edk,Palmerio:2020icb}. Then, we compared these light curves with the sensitivities of the CTAO-North array for a $5\sigma$ detection for the same energy values and assuming three different exposure times (1 min, 10 min and 60 min, see Figure \ref{fig:EBL}). This comparison can provide a good proxy to evaluate if future generation IACTs will be able to detect objects with similar VHE emission and until which time with respect to the GRB trigger time. The emission at 25 GeV is chosen because it represents the energy threshold for the best observational conditions of IACTs. However, the emission at 250 GeV reproduces a more realistic observational condition for IACTs considering external factors, i.e., high zenith observations, high night sky background, that can easily degrade the energy threshold at the level of a few hundreds of GeV.

This comparison shows that early afterglow at 25 GeV and 250 GeV can be detected up to $\sim 1 -10$ ks with an exposure from 1 to a few tens of minutes. This calculation is performed in the optimistic assumption that the GRB is localized and followed-up from IACTs within the first tens or hundreds of seconds. It is of particular interest to note that the observed emission at 25 GeV is almost unaffected by the distance since the EBL impact is low at these energies. This is evident when considering the light curves obtained for the simulated event with the same intrinsic properties of GRB\,230812B at redshift $z =2.0$. In this case, while the expected observed flux at 250 GeV is always below the CTAO-North sensitivities, the expected observed flux at 25 GeV is above the sensitivities up to $1 -10$ ks. This indicates that future generation instruments like the LSTs or the CTAO-North array will be able to sensibly expand the horizon of detection of GRBs exploiting their low energy threshold.

Given that the afterglow of GRB~230812B is estimated to be TeV-bright, the TeV detection depends on the early localization due to the limited FoV of the IACTs ($\sim$5-7 deg$^2$). In this work, we proposed the possibility of performing a systematic tiling to the localization provided by the \textit{Fermi}/GBM GCN notices distributed publicly right after the detection of the burst. The follow-up strategy should be made adaptive in such a way that the GCN-notices received within a few minutes are followed and checked for a better sky-localization. Through simple estimates, we showed that the GRB localization could have been covered and a TeV component might have been detected given the strategies mentioned above. However, the assumptions are rather simple and a development of a more realistic estimates for time delays/ tiling strategy is ongoing. This will be discussed in a future publication (Macera et al. in preparation).\par

We summarize our work as follows:

\begin{itemize}[label=\textbullet, nosep, leftmargin=*]
    \item The emission mechanism in GRBs through the detection of multiwavelength afterglow has been explored in details through GRB~230812B thanks to the exceptional brightness and the GRB being visible (inside the Fermi FoV, both GBM and LAT) up to about 1000\,s.\\
    \item Although detection of the GeV emission in GRBs is common, due to the limited sensitivity in GeV band it is challenging to establish the GeV emission to originate from the synchrotron and/or inverse Compton of the SSC model. We extracted the early afterglow emission by MeV detectors. 
    By combining the sub-MeV and GeV early afterglow data, we constructed a multiwavelength spectrum. This spectrum was then reproduced with a synchrotron and SSC scenario.
    \\
    \item We estimated the emission for gamma-rays in the lower energy band (25\,GeV), which is achieved under optimistic observing conditions, and in the higher energy band (250 GeV), which represents the threshold under realistic observing conditions. The early afterglow expected emission at 25 GeV is detectable with CTAO-North also up to redshift $z = 2$ until a few hours from the trigger time. The emission at hundreds of GeV is detectable for GRB~230812B until a few tens of minutes (10$^{3}$\,s). However, intrinsic emission is highly attenuated at higher redshift (z=2) even at energies around hundreds of GeV.\\
    \item Although we state that sub-TeV emission from GRB~230812B might have been potentially detectable by IACT, the main challenge is to design an optimal observational strategy. The TeV bright GRBs detected so far are triggered and localized by \textit{Swift}/BAT. This is due to the availability of a precise position in the sky. Instead, the precise localization of GRB~230812B was not available until 30\,ks. We demonstrated that optimizing the observational strategy could enhance the probability of detecting GRBs with CTA/LST. This involves promptly responding to early alerts from MeV detectors by targeting the initial, broader sky localizations provided in GCN notices. Tiling observations within the localization region should be performed, with real-time updates as more precise localizations become available.

\begin{acknowledgement}
We acknowledge the use of data from the uGMRT for this study.
We thank the staff of the uGMRT that made these observations possible. uGMRT is run by the National Centre for Radio Astrophysics of the Tata Institute of Fundamental Research. SM carried out a part of the research at GSSI, which was funded by the Fendi Prize money awarded to M. Branchesi. We thank Dr. V. Chitnis for valuable discussions and suggestions during the preparation of the uGMRT proposal. We also thank Dr. S. Mangla for their support in uGMRT data analysis. We thank the Department of Science and Technology (DST), India, for financial support under grant number CRG/2022/009332. We acknowledge the use of public data from \textit{Fermi} and \textit{Swift} missions and thank the respective teams for making the data available. BB and MB acknowledge financial support from the Italian Ministry of University and Research (MUR) for the PRIN grant METE under contract no. 2020KB33TP. DM acknowledges “funding by the European UnionNextGenerationEU” RFF M4C2 project IR0000012 CTA+.
LN acknowledges funding  by the European Union-Next Generation EU, PRIN 2022 RFF M4C21.1 (202298J7KT - PEACE).
\end{acknowledgement}

\end{itemize}

\bibliography{main.bib}{}
\bibliographystyle{aa}

\end{document}